\documentclass[prl,twocolumn,epsf,psfig]{revtex4}
\usepackage{graphicx}
\DeclareGraphicsExtensions{.pdf,.png,.gif,.jpg}
\usepackage{float}
\usepackage{subfig}
\usepackage{epsfig}
\usepackage{wrapfig}
\usepackage{bbold}
\usepackage{amsmath}
\usepackage{color}
\usepackage{xcolor}

\restylefloat{figure}

\begin{document}

\title{Quantum Filtering of Hydrogen Isotopes through Graphene}

\author{Joshua Hale and Theja N. De Silva\footnote{tdesilva@augusta.edu}}
\affiliation{Department of Physics and Biophysics, Augusta University, Augusta, Georgia 30912, USA.}

\begin{abstract}
Driven by the growing demand in the energy, medical, and industrial sectors, we investigate a hydrogen isotope separation technique that offers both a high separation factor and economic feasibility. Our findings reveal that filtering isotopes through two-dimensional graphene layers provides an exceptionally efficient quantum-mechanical method for isotope separation. Using a recently developed analytical pairwise potential between hydrogen isotopes and carbon atoms in graphene, we examine the classical trajectories of isotopes near the graphene layer, as well as the quantum-mechanical tunneling properties of isotopes through the graphene layer. Using various quantum-mechanical methods, we calculate both the isotope tunneling probabilities and the quantum-mechanical isotope sticking probabilities. Our study shows that quantum filtering through graphene layers can be an effective technique for enriching deuterium by separating it from protium.

\end{abstract}

\maketitle

\section{I. Introduction}

Heavy hydrogen isotopes have a wide range of applications across various industries, including the medical and nuclear energy sectors. In medicine, these isotopes are used in isotope-based tracing techniques~\cite{1} and cancer therapies~\cite{2}. In the energy sector, they serve as fuel for fusion reactors, nuclear power generation, which produces significantly less radiation than other energy sources~\cite{3}, presenting a promising solution for meeting sustainable energy demands. Among the methods of generating nuclear power, nuclear fusion is expected to offer numerous advantages over nuclear fission and nuclear decay, including enhanced safety, reduced radiation, and minimal nuclear waste~\cite{4, 5, 6, 7}. In addition, heavy water (D$_2$O) plays a crucial role in various industrial applications, particularly in nuclear engineering.

Unfortunately, the natural abundance of deuterium, which consists of a proton and a neutron, is only about 0.013\% in surface water~\cite{8}. Furthermore, separating hydrogen isotopes from a mixture is particularly challenging because of their similar physicochemical properties. Traditional separation methods are energy-intensive and typically achieve separation factors of less than 2.5~\cite{9, 10}. As a result, the development of alternative, low-energy, high-efficiency isotope separation techniques is crucial for advancing both engineering and medical technologies in the future.

Current commercial hydrogen isotope separation methods~\cite{11}, such as water distillation~\cite{12} and water-chemical exchange techniques~\cite{13}, are highly energy intensive and produce relatively low separation factors (the ratio of component concentrations). Water distillation relies on the small difference in boiling points (1.4°C) between water and heavy water, while chemical exchange methods involve isotope exchange between water and other chemical compounds. The challenges posed by low separation factors and high energy consumption in these conventional techniques have sparked the need for more efficient and alternative approaches to isotope separation.

One promising area of current research is the selective sieving of quantum particles through nanoporous materials~\cite{14}. The nuclear quantum effect enables the selective separation of atomic species through porous materials. When the de Broglie wavelength of a particle is comparable to the pore diameter, the particle can penetrate the material. For example, at room temperature, the de Broglie wavelengths of hydrogen and helium are $1.03 A^\circ$  and $0.73 A^\circ$, respectively. As a result, hydrogen can pass through materials with smaller pore sizes, allowing for effective sieving.

Unlike conventional membranes used for atomic sieving, van der Waals layered materials can be used to effectively separate hydrogen isotopes~\cite{15}. Monolayer graphene is one of the widely studied materials for the selective permeation of hydrogen isotopes~\cite{16, 17, 18, 19}. Graphene, due to its unique structure and properties, has shown potential for efficient isotope separation. Unlike conventional methods, which are energy-intensive and yield low separation factors, graphene-based membranes could offer a more effective and energy-efficient approach. The atomic-scale thickness, unique two-dimensional structure, and high permeability of graphene make it an ideal candidate for selectively quantum filtering hydrogen isotopes, providing a potential breakthrough in isotope separation technologies. Quantum effects, particularly tunneling, play a crucial role in the filtering process. Owing to their exceptional thermal, chemical, and mechanical stability, honeycomb-like structured two-dimensional graphene and hexagonal boron nitride (h-BN) have been used in the center of intensive research for nano-scale electronic applications~\cite{20, 21, 22, 23, 24, 25, 26, 27, 28, 29, 30, 31, 32, 33, 34, 35, 36}. However, limited research has been conducted on the barrier properties, diffusion, and tunneling behaviors of graphene and h-BN layers. Due to the difference in de Broglie wavelengths of hydrogen isotopes and their comparability to the pore sizes of graphene and h-BN, hydrogen isotopes interacting with these layered materials can exhibit a wide range of quantum transport phenomena. The difference in de Broglie wavelengths allows hydrogen isotopes to selectively experience quantum confinement, tunneling, and diffusion effects. Recent theoretical studies on the dynamic behaviors of lithium ions on graphene surfaces, including diffusion energy, equilibrium positions, and migration trajectories, have employed first-principles density functional theory and molecular dynamics simulations~\cite{37, 38,39,40, 41, 42}. In addition, various analytical methods have been used to explore the mechanical properties of atoms on layered surfaces~\cite{43, 44, 45}. On the experimental side, the mechanical properties of the atoms on these surfaces have been investigated through scanning tunneling microscope measurements and other techniques~\cite{46, 47}. However, all of these studies focus on atom/ion migration parallel to the surface, and the quantum diffusion and tunneling of hydrogen isotopes through the layers are yet to be explored. Furthermore, limited theoretical and experimental studies on atomic selectivity at room temperature through vacancy-free graphene and h-BN monolayer are controversial. Density functional theory-based calculations show that the energy barrier for proton transfer through pristine graphene sheets at ambient conditions is approximately $1.4 ev$ to $1.6 eV$~\cite{48, 49, 50}, suggesting that pristine graphene is impermeable to protons under these conditions. However, experimental measurements indicate that the energy barrier for protons passing through vacancy-free graphene is around $1.0 eV$, which is about $0.5 eV$ lower than the theoretical predictions~\cite{17, 51, 52}. This discrepancy suggests that protons exhibit high permeability and isotope selectivity in pristine graphene sheets. Proposed explanations for the lower energy barrier, such as atomic defects~\cite{53} or local hydrogenation~\cite{18, 54}, fail to account for both the high permeability and atomic selectivity observed in monolayer graphene.

Our theoretical study in this paper is mainly motivated by a few recent experiments showing unexpectedly high permeability for proton and deuteron through monolayer graphene under ambient conditions with high selectivity at room temperature~\cite{17, 55, 56, 57}. This study explores the quantum-mechanical tunneling phenomena of hydrogen isotopes through two-dimensional monolayer graphene. In order to better understand the sieving and diffusion mechanisms of hydrogen isotopes through graphene, we calculate both the classical trajectories and the quantum-mechanical tunneling and sticking probabilities. These calculations are performed by modeling the graphene energy barrier for isotopes using a recently developed continuous analytical model, along with various quantum mechanical techniques.

This paper is organized as follows. In Section II, we examine the interaction potential energy of an isotope atom near an ideal, infinite, defect-free graphene monolayer. Section III focuses on the classical dynamics of isotopes approaching the graphene surface, where we compute their trajectories by numerically solving Newton’s laws of motion. In Section IV, we explore the quantum mechanical tunneling behavior of isotopes through the graphene layer, employing a series of approximate theoretical methods to estimate tunneling probabilities. Section V investigates the quantum adsorption characteristics of isotopes, assuming they can adhere to the graphene surface by forming a bound state and exciting a lattice phonon. Finally, in Section VI, we summarize our main findings and conclusions.

\section{II. Isotope potential energy near the graphene surface}

In polymer-based electrolyte electrochemical hydrogen pumping, graphene is used as a permeation electrode at the cathode. When a controlled bias is applied between the electrodes, a mixture of H$_2$ and D$_2$ gas fed to the anode is electrochemically oxidized to protons and deuterons. These isotopes are passed through a proton exchange membrane known as Nafion, and protons selectively permeate the graphene at the cathode~\cite{17, 55}. Notice that the hydrons (protons and deuterons) are in their ionic forms ($H^{+}$ and $D^{+}$) as they diffuse through Nafion. As proposed by other groups~\cite{16, 18, 58, 59, 60, 61, 62}, we believe that the mechanism of isotope separation in graphene is a quantum-tunneling effect.

An elegant analytical expression for the potential energy between an infinite graphene surface and a host atom near the surface was derived in Ref.~\cite{43}. This is done by using a 6-12 Leonard-Jones potential between the host atom near the graphene surface and the i-th carbon atom on the graphene, in the form,

\begin{eqnarray}
u_J(\vec{r}_{iJ}) = 4 \epsilon_J \biggr[\biggr (\frac{\sigma_J}{r_{iJ}} \biggr)^{12} - \biggr (\frac{\sigma_J}{r_{iJ}} \biggr)^{6} \biggr],
\end{eqnarray}

\noindent where $\sigma_{J}$ denotes the zero-crossing distance of the potential for the isotope $J =$ \text{H}(proton) or $J=$ \text{D} (deuteron). Here, $\epsilon_J$ is the depth of the potential well for the J-isotope, and $r_{iJ}$ is the distance between the $J$-isotope atom and the i-th carbon atom in graphene. Assuming that the monolayer graphene is a perfect, infinitely large plane, the periodic nature of the honeycomb lattice structure allows Ref.~\cite{43} to derive the total potential energy between an isotope and the graphene layer as:

\begin{eqnarray}
U_J = \frac{u_J}{\epsilon_H} = 4 \pi I_J \biggr(\frac{2 \sigma_J^{12}}{5 z^{10}} - \frac{\sigma_J^6}{z^4} \biggr) - [\alpha_J(z) - \beta_J(z)] \\ \nonumber \times \biggr[2 \cos \biggr(\frac{2 \pi y}{\sqrt{3}}\biggr) \cos \biggr(2 \pi x \biggr) + \cos \biggr(\frac{4 \pi y}{\sqrt{3}} \biggr)\biggr].
\end{eqnarray}

\begin{figure}
\includegraphics[width=0.9 \columnwidth]{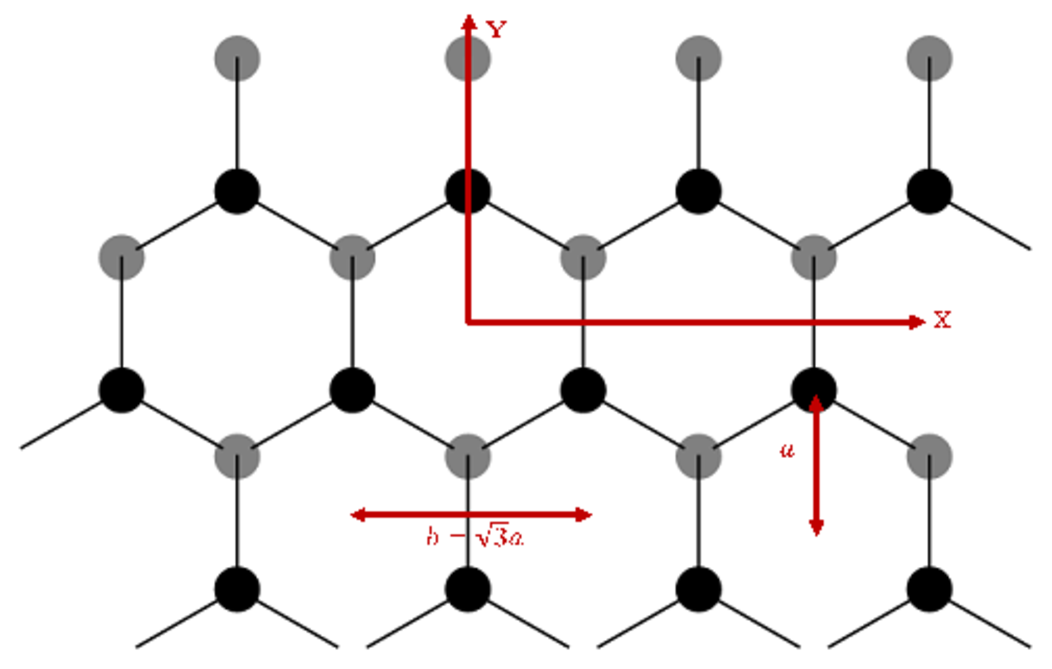}
\caption{Schematic view of the graphene sheet at $z = 0$. Notice the Cartesian coordinate system where the $x$-axis and $y$-axis are lying in the infinite graphene layer. The hydrogen isotope atom is migrating from a point where the Cartesian coordinate $(x,y,z)$ with $z > 0$ is located above the graphene plane. Notice $b = \sqrt{3} a$, the distance between the centers of two hexagons in the honeycomb lattice is set as our unit of length.}\label{graphene}
\end{figure}

\noindent Notice that the dimensionless potential energy $U_J$ is scaled with the depth of the LJ potential well $\epsilon_H$. To the best of our knowledge, specific Lennard–Jones parameters for graphene–proton ($H^{+}$) and graphene–deuteron ($D^{+}$) interactions are not available. Therefore, in our calculations, we use the known LJ parameters for graphene–atomic hydrogen (H) as an approximation for graphene–proton ($H^{+}$). The LJ parameters for C-H and C-D (neutral atoms) are generally considered identical. However, for $C-H^{+}$ and $C-D^{+}$, slight differences are expected because the nuclei lack surrounding electron clouds. In this charged–neutral system, the dominant interaction is the electrostatic attraction to the induced polarization of the carbon, rather than short-range Pauli repulsion. Moreover, quantum nuclear effects such as zero-point energy and isotope shifts modify the effective interaction potential. These effects typically make the LJ well depth for the heavier isotope ($D^{+}$) slightly deeper, implying stronger binding than for $H^{+}$. Based on sample calculations including zero-point and quantum nuclear corrections, we estimate that the effective LJ well depths for $H^{+}$ and $D^{+}$ differ by about $5\%$~\cite{62AA, 62BB}. While individual tunneling probabilities for each isotope depend on the specific LJ parameters, their ratio is primarily governed by the mass difference and the relative difference in LJ well depths. Our qualitative conclusions remain robust as long as the $5\%$ isotope-dependent difference in well depth is maintained, regardless of the precise LJ parameters used. Therefore, we approximate $C-H^{+}$ potential well depth by $C-H$ potential well depth $\epsilon_H\simeq 3.09 \times 10^{-25}$J~\cite{63}. The value of $\epsilon_D$ is also unknown, so we approximate it by using $\epsilon_H = 95 \% \epsilon_D$. Thus, the parameter $I_J$ in the potential energy, $I_H = 1.0$ and $I_D = 1.05$. Further, all lengths are scaled relative to the distance between the centers of two hexagons in the honeycomb lattice, where $b = \sqrt{3} a$ with honeycomb lattice constant $a = 1.42 A^\circ$ (see FIG.~\ref{graphene} for details). Using $\sigma_H = 1.2028$ [63], we set $\sigma_D = 1.2148$, which is approximately $1.01 \%$ larger than $ \sigma_H$. The estimates for $\epsilon_D$ and $\sigma_D$ are reasonable, as the slightly higher mass of the deuterium results in slightly larger Lennard-Jones parameters for the $C-D$ interactions (These estimates are determined based on the L-J interaction parameters for $H_2$ and $D_2$). As mentioned above, in realistic experimental setups, graphene is typically embedded within a polymer electrolyte membrane, with hydrons transported through a medium such as Nafion. Under these conditions, the interaction potential between the isotopes and graphene may differ slightly from that in vacuum. However, because our focus is on the ratio of isotope tunneling probabilities, we expect the inclusion of medium effects to have only a negligible impact on our results. Above, $(x,y,z)$ represents the scaled Cartesian coordinates of the J-isotope positioned above the graphene sheet at $z=0$, as illustrated in the FIG. \ref{graphene}. The two functions $\alpha_J(z)$ and $\beta_J(z)$ in the scaled interaction potential originated from summing over all carbon atoms at the position $\vec{r}_{iJ}$ on the graphene sheet are given by,

\begin{eqnarray}
\alpha_J(z) = \frac{\sigma_J^{12}}{30} \biggr(\frac{4 \pi}{2z} \biggr)^5 K_5(4 \pi z),
\end{eqnarray}

\noindent and

\begin{eqnarray}
\beta_J(z) = 2 \sigma_J^6 \biggr(\frac{4 \pi}{2z} \biggr)^2 K_2(4 \pi z),
\end{eqnarray}

\begin{figure}
\includegraphics[width=0.9 \columnwidth]{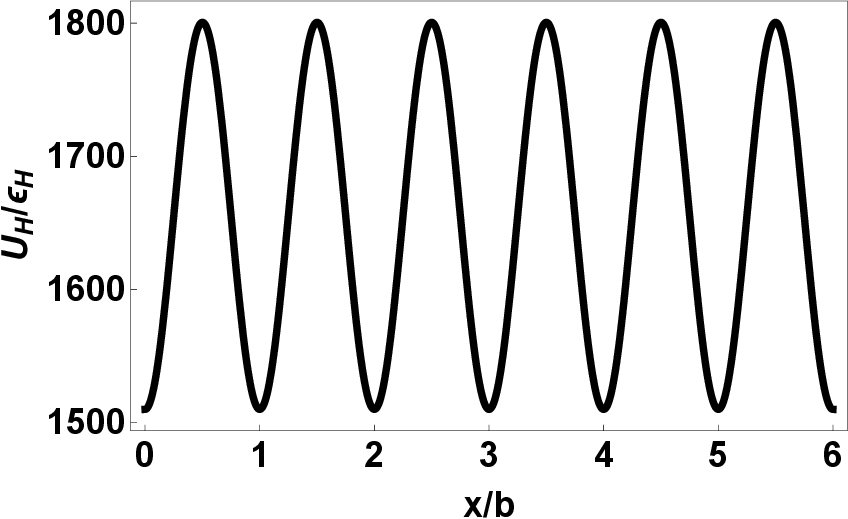}
\caption{The variation of the interaction potential energy for an H-isotope at a distance $a$ above the graphene surface along the $x$-axis.}\label{PExH}
\end{figure}

\begin{figure}
\includegraphics[width=0.9 \columnwidth]{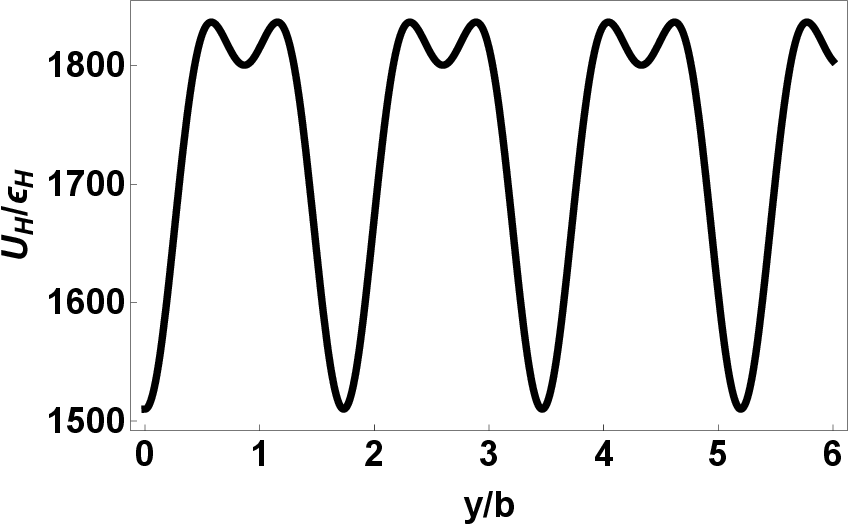}
\caption{The variation of the interaction potential energy for an H-isotope at a distance of $a$ above the graphene surface along the $y$-axis. }\label{PEyH}
\end{figure}

\begin{figure}
\includegraphics[width=0.9 \columnwidth]{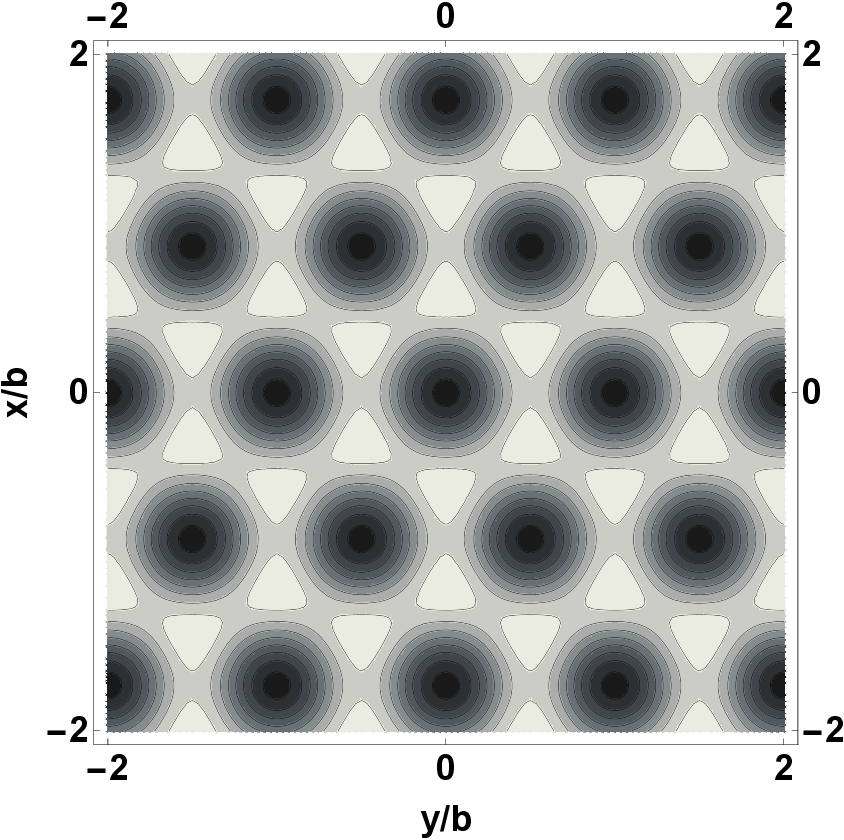}
\caption{Contour plots showing the variation of the interaction potential energy for H-isotopes at a distance of $1.1a$ above the graphene surface on the $xy$-plane. The color gradient, ranging from black to white, indicates an increase in the magnitude of the potential.}\label{PECP}
\end{figure}

\noindent where $K_n(z)$ is the $n$th-order modified Bessel function and the $4 \pi$ in these equations is the scaled magnitude of the reciprocal lattice vector~\cite{43}. The spatial variation of the potential energy on H-isotope is shown in FIGS. \ref{PExH}, \ref{PEyH}, and \ref{PECP}. The interaction potential energy of the D-isotope exhibits the same qualitative spatial behavior.

\begin{figure}
\includegraphics[width=0.9 \columnwidth]{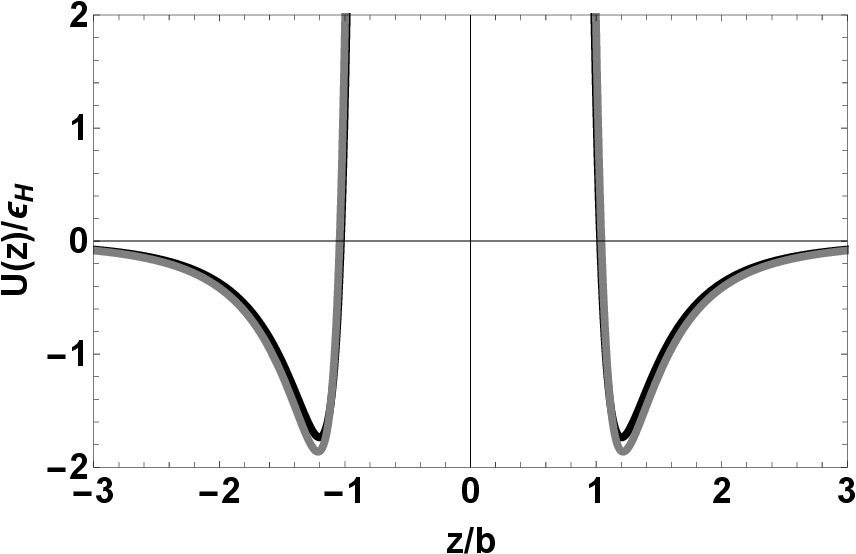}
\caption{The variation of the interaction potential energy for isotopes along the $z$-axis. The figure shows both the attractive part at large $z$ and the repulsive part at small $z$ for the H-isotope (black) and D-isotope (gray).}\label{PEzS}
\end{figure}

\begin{figure}
\includegraphics[width=0.9 \columnwidth]{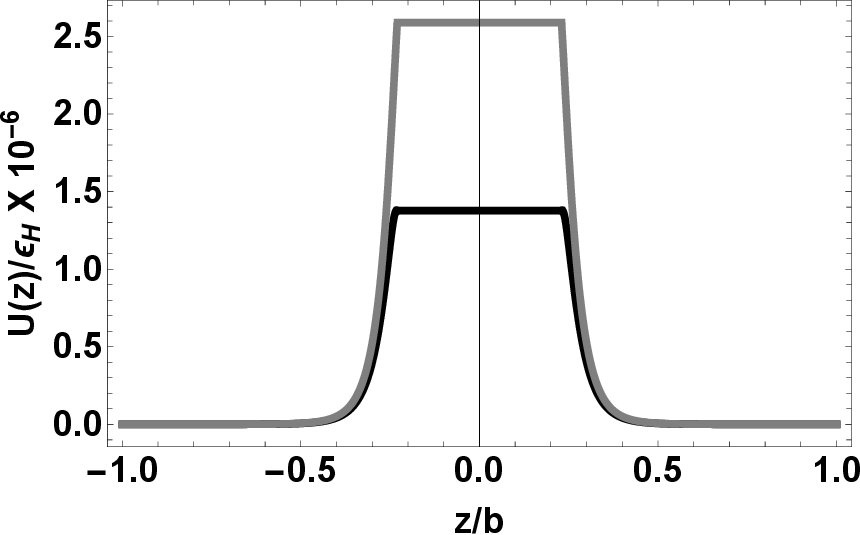}
\caption{The variation of interaction potential energy for isotopes close to the graphene surface. The figure displays the small-$z$ repulsive part for both H-isotope (black) and D-isotope (gray). The potential energy functional breaks down in the vicinity of the surface at $z = 0.23 b$ for the H-isotope and $z = 0.21 b$ for the D-isotope. Thus, we approximate it by flat-top potential beyond close to the graphene surface. See text for details.}\label{PEzL}
\end{figure}

\noindent The interaction potential energy variation along the  $z$-axis is shown in FIG. \ref{PEzS} and FIG. \ref{PEzL}. The potential energy functional $U(z)$ starts to decrease beyond $z=0.23 b$ for the H-isotope and $z=0.21 b$ for the D-isotope as the z-coordinate decreases. In other words, the our potential energy functional breaks down very close to the graphene surface, specifically within range $-z_{0J} \le z \le z_{0J}$. Therefore, we approximate $U_H(z) = U(z_{0H} =0.23 b)$ and $U_D(z) = U(z_{0D} = 0.21 b)$ for the range $-z_{0J} \le z \le z_{0J}$.

\section{III. Classical trajectories of the hydrogen isotopes}

\begin{figure}
\includegraphics[width=0.9 \columnwidth]{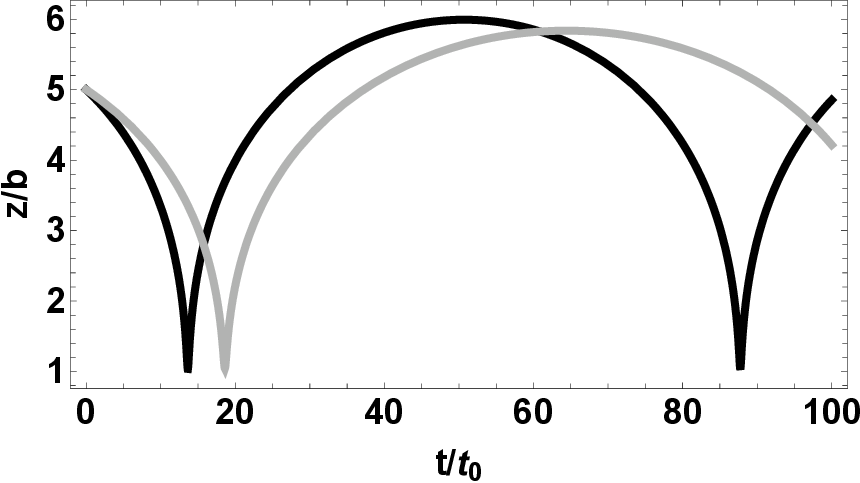}
\caption{The trajectory of the H-isotope (black) and the D-isotope (gray) perpendicular to the graphene sheet as a function of scaled time, where time is scaled with $t_0 = b \sqrt{m_H}{\epsilon_H}$.}\label{CTJ}
\end{figure}

Using the interaction potential energy functional $U_J$, we calculate the classical trajectories of the isotopes. By finding the force acting on the migrating isotope, $\vec{F} = -\vec{\nabla} U(\vec{r})$, we numerically solve the Newton's laws of motion for the trajectories of each isotope. Figure. \ref{CTJ} shows the $z$-component of the trajectory as a function of scaled time $t_0$, where the time $t$ is scaled with $ t_0 =b\sqrt{m_H/\epsilon_H}$. For the demonstration in FIG. \ref{CTJ}, we use initial coordinates for the isotopes as $(2a, 2a, 5b)$ with an initial small velocity in the $z$-direction for each isotope such that they have the same kinetic energy. The difference in trajectory results from the slight difference in the interaction potential, as well as the difference in mass, $m_H$, and $m_D = 2 m_H$. As expected, the oscillatory behavior is due to the long-distance attractive and short-distance repulsive nature of the interaction potential.

\section{IV. Quantum tunneling of the hydrogen isotopes through graphene}

To gain insight into the selective hydrogen isotope separation mechanism through a monolayer graphene layer, we apply three distinct approximate methods to calculate the quantum mechanical tunneling probabilities, arranged in increasing order of accuracy. For an infinitely large graphene layer, we can model the transport properties as one-dimensional along the $z$-direction. First, we approximate the interaction potential as an infinitely large local potential at $z =0$, and calculate the tunneling probabilities using a $\delta$-function interaction.  Next, we neglect the long-distance attractive part component of the potential and approximate it as a rectangular potential barrier, bounded between $-z_{0J} \le z \le z_{0J}$. Finally, we use Wentzel-Kramers-Broullin (WKB) approximation to compare the tunneling probabilities.

\subsection{Quantum tunneling through a $\delta$-function potential}

\begin{figure}
\includegraphics[width=0.9 \columnwidth]{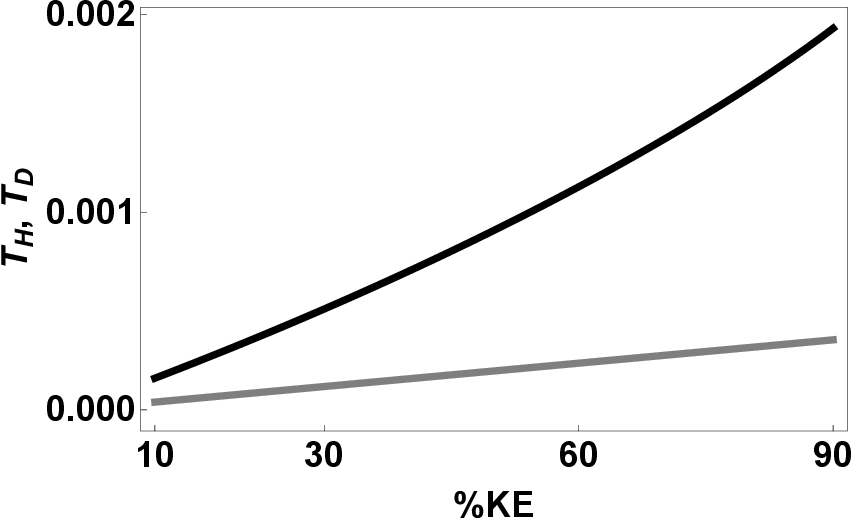}
\caption{The transmission probability of protium (black) and deuterium (deuterium) as a function of isotope kinetic energy. The interaction potential between isotopes and the graphene plane is assumed to be $\delta$-function type potential with zero width.}\label{TDI}
\end{figure}

\begin{figure}
\includegraphics[width=0.9 \columnwidth]{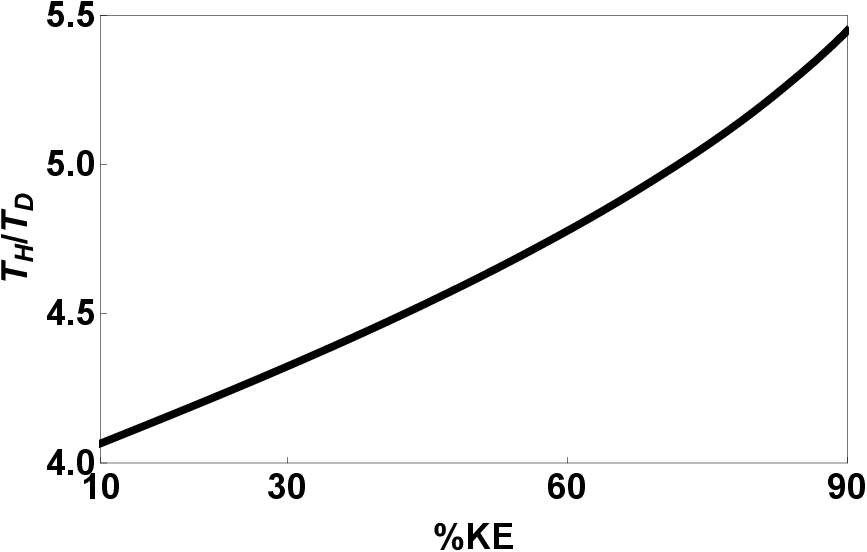}
\caption{The ratio of the tunneling transmission probability for protium and deuterium as a function of isotope kinetic energy. The interaction potential between isotopes and the graphene plane is assumed to be $\delta$-function type potential with zero width.}\label{TD}
\end{figure}

The interaction potential at the graphene surface $U_J(z_{0J})$ is much larger than the interaction potential at the position of the potential minimum, $|U_J(z_{min})|$, where $z_{min}$ is the $z$-coordinate of the most negative potential. Therefore, as the first approximation, we assume that the interaction potential takes the form $U_J(z) = \eta_J \delta (z)$, where the dimensionless parameter $\eta_J = \int_{-z_J}^{z_J} U_J(z) dz$ is scaled with $\epsilon_H b$. The integration bound $z_J$ depends on the kinetic energy, $E_J$ of the isotope at the electrolyte electrochemical hydrogen pumping anode. It is determined by the condition  $U_J(z_J)-E_J = 0$. The transmission coefficient, $T_J$, for this infinite potential with zero width can be easily calculated using standard quantum mechanical methods. The transmission coefficient, or transmission probability, in dimensionless units takes the following form:

\begin{eqnarray}
T_J = \frac{1}{1+\frac{\eta_J^2}{4 E_J} B_J},
\end{eqnarray}

\noindent where the dimensionless parameter $B_J = (2m_J/\hbar^2) b^2 \epsilon_J$. For the system under consideration, we have $B_H = 0.00562$ and $B_D = 0.01183$. In FIGS. \ref{TDI} and \ref{TD},  we plot the transition coefficients and the ratio of the tunneling transmission coefficient as a function of the percent kinetic energy of the incoming isotopes. The percent kinetic energy ($\%KE$) is defined as the $\% KE = (E_J/U_H(z_{0H})) \times 100$. In experiments, electrochemical oxidation of a mixture of H$_2$ and D$_2$ gas produces protons and deuterons, which possess the same effective charge and electric potential at the anode. As the electric potential energy is converted into kinetic energy, both isotopes acquire the same kinetic energy at the anode. Thus, we assume that both isotopes have the same kinetic energy in our calculations.

Note that the quantum transmission coefficients for the isotopes in this $\delta$-function approximation are small, on the order of $10^{-3}$, yet the ratio remains relatively large. As shown in Fig. \ref{TD}, even with this simple $\delta$-function approximation for the interaction potential between hydrogen isotopes and the graphene layer, the quantum mechanical transmission properties demonstrate that hydrogen isotopes can be separated with a separation factor greater than four for the entire isotope kinetic energy range.

\subsection{Quantum tunneling through a rectangular potential barrier}

In our second approximation, we neglect the attractive component of the interaction potential and treat the repulsive part as a rectangular potential barrier, given by $U_J(z) = U_{J0}$ for the range $-z_{0J} < z < z_{0J}$ and zero, otherwise. For the system under consideration, we have $z_{0H} = 0.23 b$, $z_{0D} = 0.21 b$ and $U_{J0} = U(z_{0J})$. The transmission coefficient for this simple potential can be easily calculated by solving the standard Schrodinger equation as part of the quantum mechanical boundary value problem, and is given by the dimensionless form,

\begin{eqnarray}
T_J^{-1} = 1 + \frac{U_{J0}^2}{4E_J (U_{J0}-E_J)} \\ \nonumber \times \sinh^2[2 z_{0J} \sqrt{B_J (U_{J0}-E_J)}].
\end{eqnarray}

\begin{figure}
\includegraphics[width=0.9 \columnwidth]{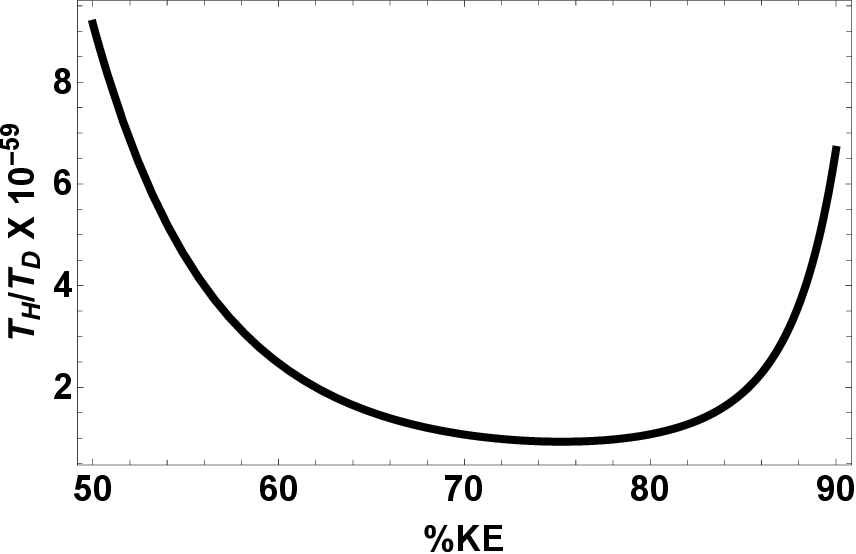}
\caption{The ratio of the tunneling transmission coefficient for protium and deuterium as a function of isotope kinetic energy. The interaction potential between isotopes and the graphene plane is assumed to be a rectangular barrier.}\label{TRB}
\end{figure}

\noindent As anticipated, we observe that the transmission coefficients for the isotopes are exponentially small, approximately $10^{-19}$ for protium and $10^{-78}$ for deuterium. Nevertheless, the ratio between them remains exceptionally large, on the order of $10^{60}$ as shown in FIG. \ref{TRB}.

\subsection{Quantum tunneling through the graphene potential barrier using WKB approximation}

In this section, we calculate the transmission coefficients for the isotopes using the well-known WKB approximation. The WKB approximation is valid only for slowly varying, wide potentials and captures the exponentially decaying part of the transmission coefficient. In this approximation, the exponentially decaying part, in dimensionless form, is given by:

\begin{eqnarray}
T_J  \propto \exp \biggr(-2 \int_{-z_{J}}^{z_{J}} k(z) dz \biggr),
\end{eqnarray}

\noindent where $k(z) = \sqrt{B_J (U_{J}(z)-E_J)}$. The correction to the WKB approximation can be included in the pre-factor, which cannot be determined within the approximation itself. However, we can determine the pre-factor using the previously discussed rectangular potential barrier calculation. By expanding the hyperbolic $\sin$ term in Eq. 6, using $\sinh(x) \simeq  e^x/2$ for large $x$, we determine the pre-factor. The final WKB transmission coefficient is then given by:

\begin{eqnarray}
T_J  = \frac{16 E_J (U_{J0}-E_J)}{U_{J0}^2} \exp \biggr(-2 \int_{-z_{J}}^{z_{J}} k(z) dz \biggr).
\end{eqnarray}

\begin{figure}
\includegraphics[width=0.9 \columnwidth]{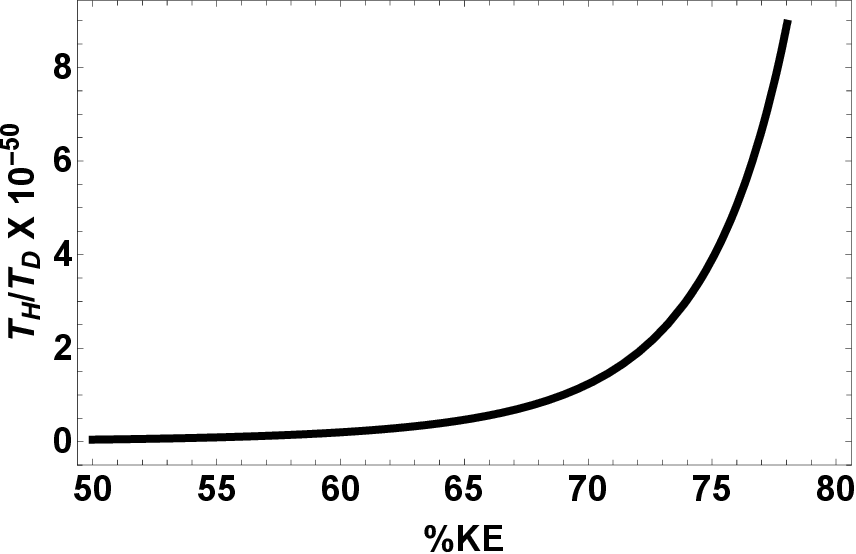}
\caption{The ratio of the tunneling transmission coefficient within WKB approximation for protium and deuterium as a function of isotope kinetic energy.}\label{TWKB}
\end{figure}

\noindent Again similar to the rectangular potential barrier approximation, the transmission coefficients within the WKB approximation for the isotopes are exponentially small, on the order of $10^{-8}$ for protium and $10^{-58}$ for deuterium. However, the ratio between them remains exceptionally large, on the order of $10^{50}$ as shown in FIG. \ref{TWKB}.

It is worth noting that, within the WKB framework, tunneling probabilities are highly sensitive to variations in the potential energy and decrease exponentially with increasing barrier height or width. However, our focus here is on isotope separation, which depends on the ratio of tunneling probabilities between isotopes. This ratio remains largely unchanged as long as the relative potential well depth is fixed as we estimated it to be about $5\%$.

\section{V. Quantum adsorption of the hydrogen isotopes on graphene}

The adsorption or quantum sticking of isotopes onto the graphene surface could further hinder tunneling through the graphene layer. Due to the isotope interactions on the graphene surface, the isotopes can be bound to the surface through phonon exchange~\cite{64}. The probability of J-isotope sticking, $S_J$, is related to the sticking rate, $R_J$, through the relation $S_J = \sqrt{2 \pi^2 m/E_J} R_J$, where $R_J$ is the transition rate, quantifying the sticking rate per unit surface area per unit incoming flux~\cite{65}. The sticking rate within Fermi's golden rule (FGR) is given by,

\begin{eqnarray}
R_J = \frac{2 \pi}{\hbar} \sum_f |\langle f|H_J|i\rangle|^2 \delta(E_f-Ei) N_q,
\end{eqnarray}

\noindent where $H_J$ is the surface-isotope coupling and $N_q = 1/(e^{\beta \hbar \omega_q}-1)$, with inverse dimensionless temperature $\beta = 1/k_BT$, is the equilibrium phonon number. The lowest-order surface-isotope coupling term, in second quantized form, is given by:

\begin{eqnarray}
H_J = -g^J_{kb} (a_k^\dagger b + b^\dagger a_k) \sum_q (p_q + p_q^\dagger),
\end{eqnarray}

\noindent where $a^\dagger_k (a_k)$ is the creation (annihilation) operator for an isotope in continuum state $|k \rangle$, $b^\dagger$ is the creation operator for an isotope bound state to the graphene plane, and $p^\dagger_q$ is the creation operator for a circularly symmetric transverse acoustic phonon in the graphene plane with energy $\hbar \omega_q$. The initial state $|i \rangle = |k \rangle \otimes|0\rangle$ is the continuum state of the isotope with zero bound isotopes and the final state $|f \rangle = |b \rangle \otimes|1\rangle$ is the isotope bound state with one phonon with momentum $\hbar q$, thus $E_f -E_i =  \hbar \omega_q -E_b - E_k$ with isotope bound state energy $E_b$. The coupling constant,

\begin{eqnarray}
g^J_{kb} = \langle k |\frac{dU_J}{dz}|b \rangle.
\end{eqnarray}

\noindent For low temperatures, where $\hbar \omega_D  \ll \beta^{-1}$ and $\omega_D$ with the Debye frequency, the phonon number can be approximated as $N_q \approx 1/(\beta \hbar \omega)$. Therefore, the finite temperature sticking rate from the FGR is~\cite{66},
\begin{eqnarray}
R_J \propto \frac{(g^J_{kb})^2}{\beta(E_k+E_b)}.
\end{eqnarray}

\noindent Using $g^J_{kb} \propto dU_J/dz$ for both isotopes, we find,

 \begin{eqnarray}
\frac{S_D}{S_H} = \sqrt{2} \alpha^2,
\end{eqnarray}

\noindent where $\alpha$ is defined as $g^H_{kb} = \alpha g^D_{kb}$. Without knowing the scattering wavefunction and the bound state wavefunction of the isotopes, it is not possible to determine the parameter $\alpha$ exactly. However, we can estimate a range by using approximate wavefunctions for the isotopes subjected to the graphene potential. Assuming normal incidence, we denote the continuum quantum states and bound state wavefunctions in real space as $|k_J \rangle \sim \sin (k_Jz+\delta_0)$ and $|b_J \rangle \sim e^{-k_{bJ} z}$, respectively,  where $k_{bJ} = \sqrt{2m E_{bJ}}/\hbar$~\cite{64, 65}. The coupling constant, expressed in terms of dimensionless parameters in the real-space representation, is then given by:

\begin{eqnarray}
g^J_{kb} \propto\int_0^\infty \sin (k_Jz+\delta_0) \frac{dU_J}{dz} e^{-k_{bJ} z} dz.
\end{eqnarray}

\noindent The integrand is oscillatory and rapidly decays due to the nature of the wavefunctions. Furthermore, $ \gamma_J = \frac{dU_J}{dz}_{z = z^\ast}$ exhibits a sharp peak in the vicinity of the potential at $z \equiv z^\ast \simeq 0.25/b$. Therefore, we approximate the integral,

\begin{eqnarray}
g^J_{kb} \propto \biggr(\frac{dU_J}{dz}\biggr)_{z = z^\ast} \int_0^\infty \sin (k_Jz+\delta_0)  e^{-k_{bJ} z} dz.
\end{eqnarray}

\noindent The integral can then be completed analytically, yielding, $\cos(k_Jz + \delta_0 +\phi)/\sqrt{k_J^2 +k_{bj}^2}$, where $\phi = \tan^{-1} (k_J/k_{Jb})$. However, in order to estimate the adsorption probability, we can assume that the kinetic and binding energies of the isotopes are the same. In this case, by canceling the integral part, we numerically estimate $\gamma_D/\gamma_H \approx 4$. This  results in the ratio of adsorption probabilities, $S_D/S_H \simeq \sqrt{2} \times 4 = 5.7$. This simple estimate indicates that deuterium has a higher tendency to adsorb onto the graphene surface compared to protium, enhancing isotope separation. However, if we relax other assumptions, such as equal binding energies etc., we believe that experiments have greater controllability of separating protium from deuterium. In our adsorption probability calculations, the isotope-carbon interaction was modeled using the LJ potential, and the adsorption ratio was estimated from wave functions corresponding to the short-range potential. However, incident charged isotopes can also experience a long-range Coulomb-like interaction and may undergo charge transfer with the graphene surface. We did not include this effect in our calculations, as our primary goal was to estimate the adsorption probability ratio. The difference between long-range and short-range interactions enters through the dependence of the sticking rate on the incident energy. For example, while the sticking rate approaches unity in the classical limit, it scales as $\sqrt{E}$ for short-range interactions and as $E$ for long-range interactions at low energies~\cite{64}. When comparing isotopes with the same incident kinetic energy, this common energy dependence cancels out in the ratio, so our qualitative estimate should remain valid regardless of whether charge transfer effects are included.

\section{VII. Conclusions and Summary}

We investigated the classical trajectories, quantum mechanical tunneling probabilities, and quantum adsorption probabilities of hydrogen isotopes in the potential field of an ideal, infinitely extended graphene layer. We employed a series of approximate quantum mechanical approaches to compute the transition probabilities of isotopes through the graphene layer. As anticipated, we find tunneling probabilities of isotopes are exponentially small due to the significant potential barrier near the graphene surface. However, we observed that the ratio of the tunneling probabilities between protium and deuterium-isotopes is exponentially large, which is attributed to both the mass difference and variations in potential parameters between the isotopes. Moreover, we find that the adsorption characteristics of isotopes contribute significantly to improved separation efficiency. Our study finds that quantum filtering can serve as an effective isotope separation technique, even though classical permeation through graphene is unlikely due to the high energy barrier predicted by DFT. Therefore, we conclude that consecutive quantum filtering through graphene layers can be a strong and effective ideal experimental technique for isotope separation.

In this work, we model the nonbonded interaction between carbon atoms and protons (H$^{+}$) using the same Lennard–Jones parameters as those commonly applied for carbon–hydrogen (C–H) interactions. Since the Lennard-Jones potential is designed to capture non-bonded van der Waals interactions, specifically a short-range repulsion and a longer-range attraction, it depends primarily on atomic size and polarizability rather than the electronic state in high detail. For a proton (H$^{+}$) in many environments, especially when it has no bound electrons, the physical size is small, and the dispersion interactions are negligible. Thus, the attractive interaction for an electron-less proton is very weak and using the C–H values for well depth will overestimate attraction. However, at close approach, the proton interacts with the electron cloud of carbon in a manner similar to neutral hydrogen, so the repulsive component is expected to be comparable. Therefore, adopting standard C–H parameters for C–H$^{+}$ provides a reasonable simplification allowing us to focus on the tunneling through the graphene layer, while keeping the empirical model consistent and computationally tractable. Further, in this work, we investigated the isotope effect on quantum tunneling probabilities as a mechanism for isotope separation. Zero-point energy is not explicitly included in our tunneling calculations. However, it was taken into account when estimating the $5\%$ difference in potential well depth between the proton and deuteron. Therefore, we expect that the explicit inclusion of zero-point energy will not lead to any significant change in our conclusions.

It is important to note that the results presented in this paper are based on a series of approximations, parameter estimations, and a truncated potential near the graphene surface. For instance, all quantum mechanical calculations were carried out using well-justified approximations. The unknown Lennard-Jones parameters for the isotopes were estimated by exploring the values for H$_2$ and D$_2$. Additionally, our potential energy functional becomes invalid in close proximity to the graphene surface, where we assume a flat-top potential instead. Despite these conservative assumptions, our findings remain highly promising. We believe these approximations are reasonable and, as such, our results demonstrate strong potential and robustness for practical applications under realistic conditions.

\section{VIII. ACKNOWLEDGMENTS}

We gratefully acknowledge the partial support provided by Augusta University’s Center for Undergraduate Research and Scholarship (CURS) through the Summer Scholars Program and travel grants.

\section{AUTHOR DECLARATIONS}

\subsection{Conflict of Interest}

The authors have no conflicts to disclose.

\subsection{Author Contributions}

\textbf{Joshua Hale:}Formal analysis (lead); Investigation (lead); Validation (equal); Visualization (equal); Writing – original draft (supporting); Writing – review and editing (supporting). \textbf{Theja De Silva:} Conceptualization (lead); Data curation (lead); Formal analysis (supporting); Investigation (supporting); Methodology (lead); Project administration (lead); Writing – original draft (lead); Writing – review and editing (lead).

\section{DATA AVAILABILITY}

The data that support the findings of this study are available from the corresponding author upon reasonable request.

\end{document}